\documentclass[twocolumn,aps,pre,floatfix,superscriptaddress]{revtex4}
\usepackage{amsmath,amssymb,eucal,graphicx,bm} 
\newcommand{\ii}{{\rm i}}
\begin{document}
\title{Jamming and Tiling in Fragmentation of Rectangles}
\author{E.~Ben-Naim}
\affiliation{Theoretical Division and Center for Nonlinear Studies,
  Los Alamos National Laboratory, Los Alamos, New Mexico 87545, USA}
\author{P.~L.~Krapivsky}
\affiliation{Department of Physics, Boston University, Boston,
  Massachusetts 02215, USA}
\begin{abstract}
We investigate a stochastic process where a rectangle breaks into
smaller rectangles through a series of horizontal and vertical
fragmentation events.  We focus on the case where both the vertical
size and the horizontal size of a rectangle are discrete variables.
Because of this constraint, the system reaches a jammed state where
all rectangles are sticks, that is, rectangles with minimal width.
Sticks are frozen as they can not break any further.  The average
number of sticks in the jammed state, $S$, grows as $S\simeq
A/\sqrt{2\pi\ln A}$ with rectangle area $A$ in the large-area limit,
and remarkably, this behavior is independent of the aspect ratio.  The
distribution of stick length has a power-law tail, and further, its
moments are characterized by a nonlinear spectrum of scaling
exponents.  We also study an asymmetric breakage process where
vertical and horizontal fragmentation events are realized with
different probabilities.  In this case, there is a phase transition
between a weakly asymmetric phase where the length distribution is
independent of system size, and a strongly asymmetric phase where this
distribution depends on system size.
\end{abstract} 

\maketitle

\section{Introduction}

Fragmentation processes where large objects break into smaller ones
underlie an ever growing number of physical and natural phenomena
\cite{bdh,bkb,zxg,kbg,fl,tfg,khp}. In particular, fragmentation occurs
in soft matter systems such as polymers \cite{mrt}, active matter
\cite{mo,gtd}, granular media \cite{hp,sas}, and brittle materials
\cite{kh,alt,ghb}.

Experimental and theoretical studies of fragmentation generally focus
on the distribution of fragment size.  Typically, this 
distribution is self-similar throughout the breakage process, and it
is characterized by a single quantity, for example, the average
fragment size \cite{af,zm,cr}. Self-similarity extends to discrete
fragments and continuous ones, one-dimensional fragments and
multi-dimensional ones \cite{krb}.  However, while the fragment size
is a fluctuating quantity throughout the breakage process, the
fragment size becomes deterministic in the final state as all
fragments have the same size.  In this sense, the final state can be 
trivial.

Recently, non-trivial final states have been reported in a
multi-dimensional fragmentation process \cite{cb,qpb,bch,ch} which
models martensitic phase transformations \cite{rss,bk,tivp}.  The
system reaches a jammed state where the two-dimensional fragments are
characterized two sizes: one size is a deterministic quantity, but the
second size is a stochastic quantity.  Here, we study this planar
fragmentation process analytically, and we present a comprehensive
statistical analysis of the jammed state.

\begin{figure}[t]
\includegraphics[width=0.45\textwidth]{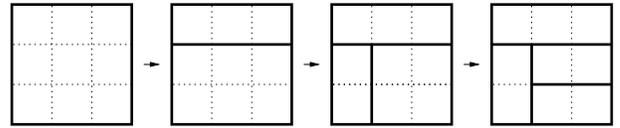}
\caption{Illustration of the fragmentation process \eqref{process}.
  Initially, the system consists of a single rectangle. Through a
  series of horizontal and vertical cuts, the system reaches a jammed
  state where fragmentation is no longer possible.}
\label{fig-ill1}
\end{figure}

We study fragmentation of rectangles with discrete horizontal and
vertical sizes (Fig.~\ref{fig-ill1}).  A rectangle can break 
vertically or horizontally into two smaller rectangles. Due to 
discreteness, rectangles with minimal vertical or horizontal size can
not break, and hence, are frozen. We refer to these frozen rectangles
as ``sticks.''  Through a sequence of random fragmentation events, the 
system which initially consists of a single rectangle, reaches a
jammed state where all rectangles are sticks (Fig.~\ref{fig-jam}). 

We find that, up to a logarithmic correction, the average number of
frozen sticks in the jammed state, $S$, grows linearly with the area 
$A$,
\begin{equation}
\label{S}
S \simeq \frac{A}{\sqrt{2\pi \ln A}}\,.
\end{equation}
Interestingly, this asymptotic behavior is universal as it applies
regardless of aspect ratio.  We also study the distribution of stick
length and find that this distribution has a power-law tail. Further,
this length distribution exhibits multi-scaling asymptotic behavior as
its moments are characterized by a nonlinear spectrum of exponents.

We also investigate an asymmetric process where horizontal and
vertical cuts are realized with different probabilities. We find a
phase transition at a critical value of the asymmetry parameter. In
the weakly asymmetric phase, the length distribution does not depend
on system size, while in the strongly asymmetric phase, this
distribution does depend on system size.

The rest of this paper is organized as follows. In Sec.~II, we
introduce the fragmentation process and develop the theoretical
techniques used throughout this investigation to obtain the leading
asymptotic behavior in the large system-size limit.  We first analyze
the average number of sticks and then consider the length distribution
of sticks. In Sec.~III, we generalize the results to the case where
horizontal and vertical fragmentation occur at different rates. Next,
in section IV, we analyze a closely related process of fragmentation
into four, rather than two, rectangles. In this case, the outcome of a
fragmentation event is deterministic, and we can also address the
total number of jammed configurations.  We conclude with a discussion
in Sec.~V. Details of several technical derivations are presented in
the Appendix.

\section{Fragmentation of Rectangles}

Initially, the system consists of a single rectangle with horizontal
size $m$, vertical size $n$, and hence, area \hbox{$A=mn$}. Both the
horizontal size and the vertical size are integer. It is convenient to
envision a square grid with \hbox{$(m-1)(n-1)$} internal grid points
embedded within the rectangle (Fig.~\ref{fig-ill1}).  In each
fragmentation event, an internal grid point is selected, and then a
cut is made along the horizontal or the vertical direction. As a
result, the rectangle breaks into two smaller ones,
\begin{equation}
\label{process}
(m,n)\to 
\begin{cases}
(i,n)+(m-i,n)& {\rm with\ prob.\ 1/2}, \\
(m,j)+(m,n-j)& {\rm with\ prob.\ 1/2}.
\end{cases}
\end{equation}
The grid point with $1\leq i\leq m-1$ and $1\leq j\leq n-1$ is chosen
at random, as is the fragmentation direction. Of course, the total
area is conserved.

\begin{figure}[t]
\includegraphics[width=0.35\textwidth,height=0.35\textwidth]{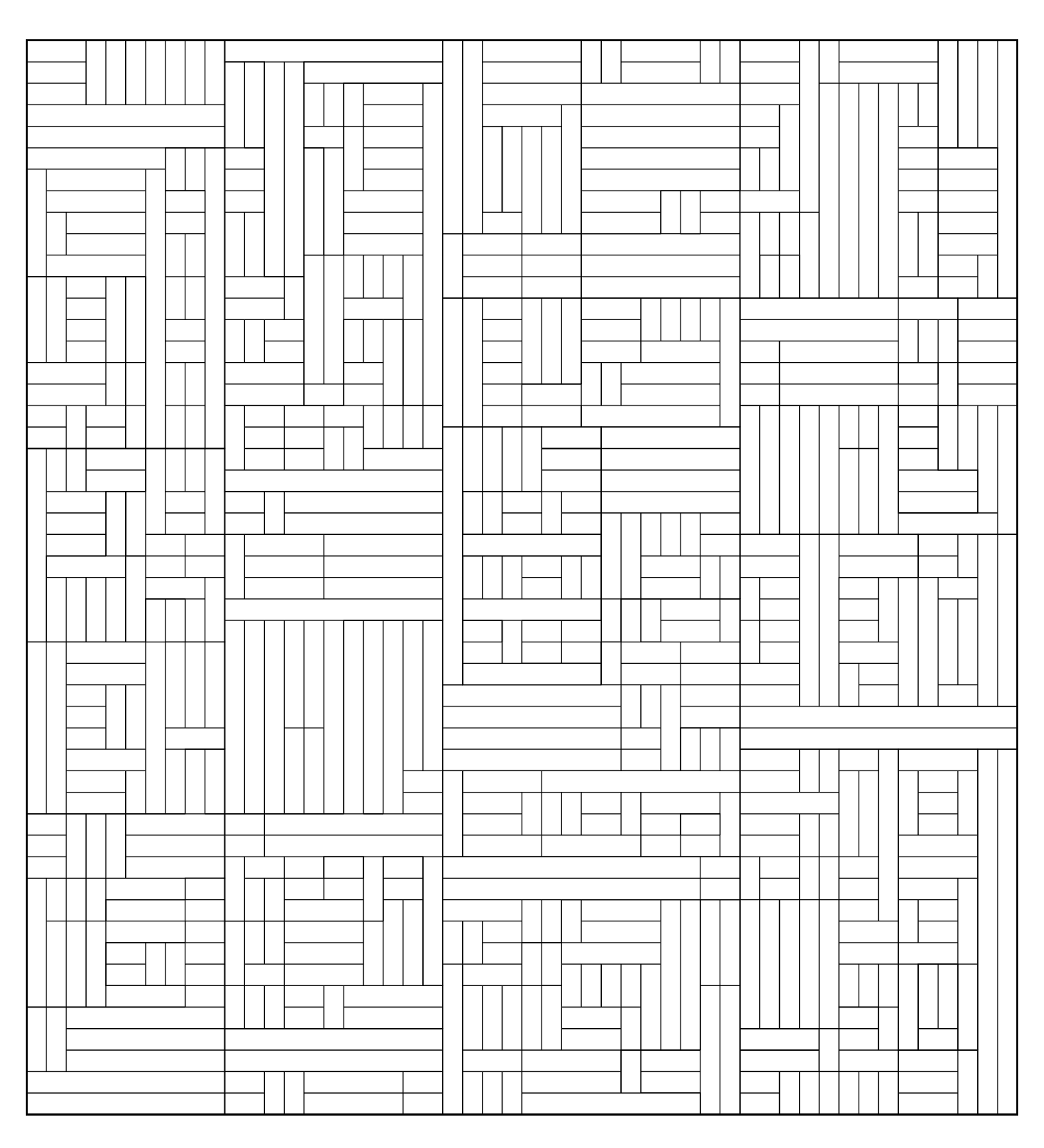}
\caption{A jammed state in a system of size $50\times 50$. The jammed
  state consists of rectangles of size $1\times k$ or $k\times 1$.}
\label{fig-jam}
\end{figure}

Fragmentation requires an internal grid point.  Therefore, rectangles
with $m>1$ and $n>1$ are active, and otherwise, rectangles with $m=1$
or $n=1$, referred to as sticks, are frozen.  The fragmentation process
\eqref{process} is repeated for every active rectangle until the
system reaches a jammed state with sticks only (Fig.~\ref{fig-jam}).
In this study, we focus on the jammed state.

Let $S(m,n)$ be the average number of sticks in the jammed state when
the initial rectangle has dimensions $m\times n$.  This average is
taken over all realizations of the random breakage process.  Since the
fragmentation process \eqref{process} is symmetric with respect to the
horizontal and the vertical direction, we expect $S(m,n)=S(n,m)$. The
average number of frozen sticks obeys the recursion 
\cite{tivp}
\begin{equation}
\label{Smn-eq}
S(m,n)=\frac{1}{m-1} \sum_{i=1}^{m-1} S(i,n)
+\frac{1}{n-1} \sum_{j=1}^{n-1}S(m,j)\,.
\end{equation}
The first term on the right-hand side accounts for the $m-1$ possible
cuts in the vertical direction, and similarly, the second term
accounts for the $n-1$ possible cuts in the horizontal direction. The
recursion equation is linear as each fragmentation event involves a
single rectangle, and it is subject to the boundary conditions
\begin{equation}
\label{Smn-bc}
S(m,1)=S(1,n)=1
\end{equation}
for all $m\geq 1$ and $n\geq 1$.

Equations \eqref{Smn-eq} and \eqref{Smn-bc} yield the average number of
frozen sticks for small rectangles, 
\begin{equation}
\label{Smn-small}
\begin{split}
&S(2,2)=2, \quad ~S(2,3)=\tfrac{5}{2}, \quad
  ~S(2,4)=\tfrac{17}{6},\\ &S(3,3)=\tfrac{7}{2}, \quad
  ~S(3,4)=\tfrac{17}{4} \quad ~S(4,4)=\tfrac{97}{18}\,.
\end{split}
\end{equation} 
Moreover, for ladders ($m=2$), 
equation \eqref{Smn-eq} simplifies to
\hbox{$S(2,n)-S(2,n-1)=\tfrac{1}{n-1}$} and therefore, 
\begin{equation}
\label{S2n}
S(2,n)= 1+H_{n-1}\,,
\end{equation}
where $H_N=\sum_{1\leq i\leq N} i^{-1}$ is the harmonic number. For
long ladders, $S(2,n)\simeq \ln n+1+\gamma$, where $\gamma = 0.57721$
is the Euler constant. It is also possible to show that as long as $m$
is finite, the leading asymptotic behavior remains logarithmic,
\begin{equation}
\label{S-large}
S(m,n)\simeq \frac{(\ln n)^{m-1}}{(m-1)!^2}\,,
\end{equation} 
in the limit $n\to\infty$.

Our main interest is the behavior for large rectangles, and
specifically, the leading asymptotics when $m\to\infty$ and
$n\to\infty$.  Hence, we treat $m$ and $n$ as continuous
variables, and replace the sums in \eqref{Smn-eq} with integrals.  The
average number of  sticks satisfies the integral equation
\begin{equation}
\label{Smn-eq1}
S(m,n) = \frac{1}{m}\int_1^m di\,S(i,n) +  \frac{1}{n}\int_1^n dj\,S(m,j)\,, 
\end{equation}
within this continuous framework.  Next, we multiply this integral
equation by the area $mn$ and then  differentiate the resulting
equation with respect to $m$ and $n$. That shows the quantity $S(m,n)$
satisfies the partial differential equation
\begin{equation}
\label{Smn-eq2}
\partial_m \partial_n \left[mnS(m,n)\right] \!=\! 
\partial_m \left[mS(m,n)\right]\!+\!\partial_n \left[nS(m,n)\right]. 
\end{equation}
Hereinafter, we use \hbox{$\partial_m = \frac{\partial}{\partial m}$}
and \hbox{$\partial_n = \frac{\partial}{\partial n}$} to denote 
partial derivatives.  Further simplification can be achieved by
introducing the logarithmic variables
\begin{equation}
\label{munu}
\mu = \ln m, \qquad \nu = \ln n\,.
\end{equation}
With this transformation, Eq.~\eqref{Smn-eq2} reduces to a partial
differential equation with constant coefficients,
\begin{equation}
\label{Smunu-eq}
\partial_\mu \partial_\nu S(\mu,\nu) = S(\mu,\nu)\,,
\end{equation}
that should be solved subject to the boundary conditions $S(\mu,0)=1$
and $S(0,\nu)=1$.

The central quantity throughout our analysis is the double Laplace
transform 
\begin{equation}
\label{Spq-def}
\widehat{S}(p,q) = 
\int_0^\infty d\mu\,e^{-p\mu} \int_0^\infty d\nu\,e^{-q\nu}\, S(\mu,\nu)\,. 
\end{equation}
It is obtained by multiplying both sides of the governing equation
\eqref{Smunu-eq} by $e^{-p\mu-q\nu}$ and then integrating over the
logarithmic variables $\mu$ and $\nu$.  By using the boundary
conditions $S(\mu,0)=1$ and $S(\nu,0)=1$, we find that double Laplace
transform is remarkably compact,
\begin{equation}
\label{Spq}
\widehat{S}(p,q) = \frac{1}{pq-1}. 
\end{equation}
Therefore, the average number of sticks in the jammed state,
$S(\mu,\nu)$, equals the inverse Laplace transform
\begin{eqnarray}
\label{Spq-int}
S(\mu,\nu) &=& \int_{-\ii \infty}^{\ii \infty} \frac{dp}{2\pi \ii}
\int_{-\ii \infty}^{\ii \infty} \frac{dq}{2\pi \ii}\,\,
\frac{e^{p\mu+q\nu}}{pq-1}\nonumber\\ 
&=&\int_{-\ii \infty}^{\ii
  \infty} \frac{dq}{2\pi \ii}\,\,\frac{1}{q}\,e^{\nu q+\mu/q}\,.
\end{eqnarray}
We perform the inversion first with respect to the conjugate variable
$p$ and then with respect to the conjugate variable $q$.  The
inversion with respect to $p$ is immediate as the integrand in the
first line has a pole at $p=q^{-1}$.

The integral over the variable $q$ in \eqref{Spq-int} has the form
\begin{equation}
\label{I-def}
I= \int_{-\ii \infty}^{\ii \infty} \frac{dq}{2\pi \ii}\,\,F(q)\,e^{\nu f(q)}\,. 
\end{equation}
The exponential dominates the integrand in the limit $\nu\to\infty$.
Further, the function $f(q)$ is maximal at the saddle point $q_*$
which is determined from $f'(q_*)=0$, and in the vicinity of this
saddle point we have
\begin{equation}
\label{saddle}
f(q)\simeq f(q_*)+\tfrac{1}{2}(q-q_*)^2 f''(q_*)\,.
\end{equation}
The integration contour in \eqref{I-def} can be along any line that
parallels the imaginary axis in the complex plane as long as
$\text{Re}(q)$ is greater than the real part of any singularity the
integrand may have. We conveniently choose a line parallel to the
imaginary axis that passes through the saddle point $q_*$.  With the
transformation of variables $q=q_*+\ii y/\sqrt{f''(q_*)}$, the
integral \eqref{I-def} reduces to a Gaussian integral.  As long as
$\text{Re}(q_*)$ exceeds the real part of the singularities of $F(q)$,
we have
\begin{equation}
\label{I-gen}
I\simeq \frac{F(q_*)\,e^{\nu f(q_*)}}{\sqrt{2\pi\nu f''(q_*)}}\,.
\end{equation}
Here, we used $\int_{-\infty}^\infty \exp(-y^2/2)dy =\sqrt{2\pi}$.

\begin{figure}[t]
\includegraphics[width=0.425\textwidth]{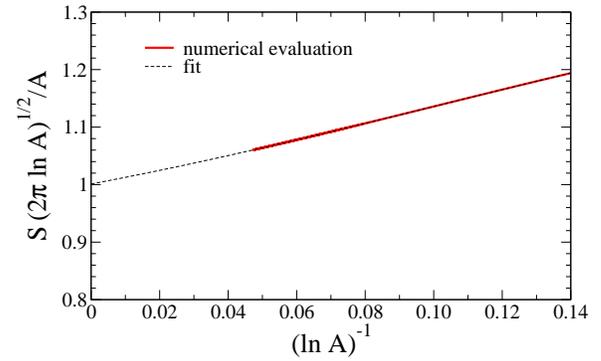}
\caption{The quantity $S\sqrt{2\pi\ln A}\,/A$ versus $(\ln
  A)^{-1}$.  The dashed line shows results of a fourth-order
  polynomial fit to the data, and the intercept agrees with the
  theoretical prediction of unity to within 0.1\%.}
\label{fig-S}
\end{figure}

First, we discuss squares, $\mu=\nu$, for which \hbox{$S(\nu,\nu)=I$}
with $I$ given in \eqref{I-def}. The quantities $F(q)=q^{-1}$ and
\begin{equation}
\label{f}
f(q)=q+q^{-1}  
\end{equation}
specify the integral \eqref{I-def}.  The saddle point is $q_*=1$, and
furthermore, \hbox{$f(q_*)=f''(q_*)=2$} and $F(q_*)=1$. By
substituting these values into the general expression \eqref{I-gen},
we obtain the leading asymptotic behavior
\begin{equation}
\label{Snunu}
S(\nu,\nu) \simeq \frac{e^{2\nu}}{\sqrt{4\pi\nu}}\,. 
\end{equation}
Equation \eqref{S} expresses this behavior in terms of the area
\hbox{$A=e^{2\nu}$}.  Results of numerical evaluation of the recursion
equation \eqref{Smn-eq} are in excellent agreement with the
theoretical prediction, and we conclude that the continuum framework 
yields exact results for the leading asymptotic behavior
(Fig.~\ref{fig-S}).

We now consider squares of arbitrary size for which $S(\mu,\nu)=I$
with $f(q)=q+(\mu/\nu)q^{-1}$ and $F(q)=q^{-1}$.  It is
straightforward to repeat the steps leading to \eqref{Snunu}, and
obtain the general behavior
\begin{equation}
\label{Smunu}
S(\mu,\nu) \simeq \frac{e^{2\sqrt{\mu\nu}}}{\sqrt{4\pi\sqrt{\mu\nu}}}\,.
\end{equation}
As expected, the average number of sticks is symmetric,
$S(\mu,\nu)=S(\nu,\mu)$, and moreover, the quantity $\nu$ in
\eqref{Snunu} is now replaced with the geometric average
$\sqrt{\mu\nu}$.  In the limit $m\to \infty$ and $n\to\infty$ with the
aspect ratio $r = \tfrac{m}{n}$ kept fixed, we have $\sqrt{\mu\nu}
\simeq (\mu+\nu)/2$.  In this limit, the leading asymptotic behavior
\eqref{Smunu} is identical to \eqref{Snunu}.  Interestingly, the
average number of jammed sticks is universal in the large-area
limit---all rectangles with the same area behave similarly. The only
requirement is that the aspect ratio is finite.  In view of this
universality, we henceforth quote results for squares without loss of
generality.

We can immediately deduce the average length of a stick in the jammed
state, $\langle k\rangle$.  Since the rectangle is covered entirely by
sticks, we have $A=S\langle k\rangle$. Therefore, the average stick
length grows logarithmically with area, $\langle k\rangle \simeq
\sqrt{2\pi \ln A}$, and this behavior is independent of the aspect ratio
in the large-area limit.

In the jammed state, the original rectangle is covered with
rectangles of size $1\times k$ or $k\times 1$ with $k\geq 2$, and we
now analyze the distribution of stick length $k$.  Let $S_k(m,n)$ be
the average number sticks of length $k$ in the jammed state, when the
initial rectangle has dimensions $m\times n$.  This quantity satisfies
two sum rules,
\begin{equation}
\label{norm}
S(m,n)=\sum_{k\geq 2} S_k(m,n)\,,\quad A=\sum_{k\geq 2} kS_k(m,n)\,.
\end{equation}
For all $k$, the quantity $S_k(m,n)$ satisfies the recursion equation
\eqref{Smn-eq}, although the boundary condition does depend on length
\begin{equation}
\label{Skmn-bc}
S_k(m,1)=\delta_{m,k}\,\qquad S_k(1,n)=\delta_{n,k}\,. 
\end{equation}
With the boundary condition $S_n(1,n)=1$, the recursion equation
\eqref{Smn-eq} implies $S_n(m,n)=1$ for all \hbox{$m<n$}, and it is
also possible to show that \hbox{$S_n(n,n)=2$}.

For large rectangles, we utilize the continuum approach once again.
As a function of the logarithmic variables defined in \eqref{munu},
the average number of frozen sticks with a given length $S_k(\mu,\nu)$
satisfies the partial differential equation \eqref{Smunu-eq}, subject
to the boundary conditions
\begin{equation}
\label{Skmunu-bc}
\begin{split}
S_k(\mu,0) &=e^{-\mu}\delta(\mu-\ln k),\\
S_k(0,\nu) &=e^{-\nu}\delta(\nu-\ln k)\,.
\end{split}
\end{equation}
To obtain these boundary conditions, we first rewrite \eqref{Skmn-bc}
as $S_k(m,1)=\delta(m-k)$ and $S_k(1,n)=\delta(n-k)$ and then perform
the transformation of variables \eqref{munu} by using \hbox{$\delta[{\cal
    F}(x)] = \delta(x-x_0)/|{\cal F}'(x_0)|$}.

Next, we repeat the steps leading to \eqref{Spq}, using the partial
differential equation \eqref{Smunu-eq} which also governs
$S_k(\mu,\nu)$ and the boundary condition \eqref{Skmunu-bc}, and
arrive at
\begin{equation}
\label{Skpq}
\widehat{S}_k(p,q)=\frac{p k^{-1-p}+q k^{-1-q}}{pq-1}\,.
\end{equation}
We note that this double Laplace transform is symmetric,
$\widehat{S}_k(p,q)=\widehat{S}_k(q,p)$ as the two terms in the
numerator are equivalent. For squares, $\mu=\nu$, it suffices to
invert only one of these terms.  We thus perform the inverse Laplace
transform of the term $q k^{-1-q}/(pq-1)$ first with respect to $p$
and then with respect to $q$, thereby leading to $S_k(\nu,\nu)=I$ with
$I$ given by \eqref{I-gen}. The integrand is specified by $F(q)=2$ and
\begin{equation}
\label{f-q}
f(q)=q+q^{-1}-(1+q)x\,,\quad x=\frac{\ln k}{\ln n}\,.
\end{equation}
The saddle point is $q_*=1/\sqrt{1-x}$, and by substituting
\hbox{$f(q_*)=2\sqrt{1-x} - x$} and \hbox{$f''(q_*) = 2(1-x)^{3/2}$}
into the general formula \eqref{I-gen}, we obtain
\begin{equation}
\label{Snunuk}
S_k(\nu,\nu)\simeq 
\frac{\exp\left[\nu\left(2\sqrt{1-x}-x\right)\right]}
{\sqrt{\pi\nu(1-x)^{3/2}}}\,.
\end{equation}

\begin{figure}[t]
\includegraphics[width=0.425\textwidth]{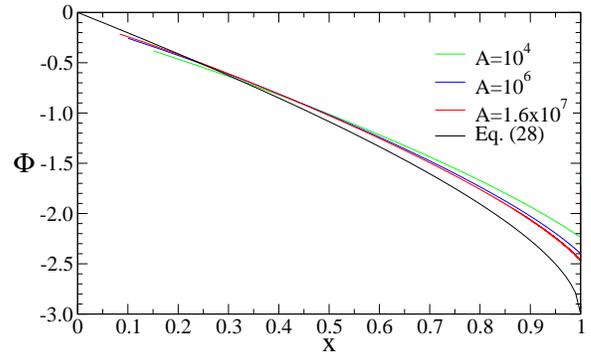}
\caption{The scaling function $\Phi(x)$ versus the scaling variable
  $x$. Results from three different systems sizes are compared with 
the  theoretical prediction.}
\label{fig-Phi}
\end{figure}

By definition, $P_k(\nu)=S_k(\nu,\nu)/S(\nu,\nu)$ is the fraction of
frozen sticks with length $k$. This distribution is normalized,
$\sum_{k\geq 2} P_k=1$, and its first moment equals the average
length, $\langle k\rangle = \sum_{k\geq 2} kP_k$.  By using equations
\eqref{Snunu} and \eqref{Snunuk} we find that the length distribution
adheres to the scaling form
\begin{equation}
\label{Pk-scaling}
\frac{\ln \left(\tfrac{1}{2}P_k\right)}{\ln n}
\simeq \Phi\left(\frac{\ln k}{\ln n}\right) 
\end{equation}
with the scaling function
\begin{equation}
\label{Phi}
\Phi(x)=2\left(\sqrt{1-x}-1\right)-x\,.
\end{equation}
Equation \eqref{Pk-scaling} constitutes an unusual scaling form as the
scaled logarithm of the length distribution $P_k$ is a universal
function of the scaled logarithm of the length $k$. As a result, the
convergence toward the ultimate asymptotic behavior is extremely slow
as it involves logarithm of system size (Fig.~\ref{fig-Phi}).
 
The scaling behavior \eqref{Pk-scaling}-\eqref{Phi} describes the
distribution at large length scales, that is, \hbox{$\ln k={\cal
    O}(\ln n)$}. Still, the small-$x$ behavior
\hbox{$\Phi(x)\simeq-2x-\tfrac{1}{4}x^2$} yields the behavior at
smaller length scales,
\begin{equation}
\label{Pk-tail}
P_k\simeq 2k^{-2}\,\exp\left[-\frac{(\ln k)^2}{4\ln n}\right]\,.
\end{equation}
Therefore, the length distribution decays as a power-law, $P_k\simeq
2k^{-2}$ \cite{tivp}, at sufficiently small length scales, \hbox{$\ln
  k \ll \sqrt{\ln n}$}. Beyond this length scale, the power-law tail
is suppressed by a log-normal term.  We also note that log-normal
distributions naturally arise in multiplicative random processes
\cite{lsa,mm}, and that the fragmentation process \eqref{process} can
be formulated as such.

\begin{figure}[t]
\includegraphics[width=0.425\textwidth]{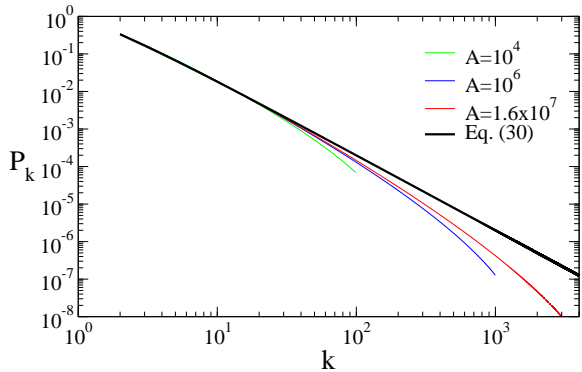}
\caption{The length distribution $P_k$ versus $k$.  Numerical results
  for three systems sizes are compared with the theoretical
  prediction.}
\label{fig-Pk}
\end{figure}

In a finite system, the power-law tail holds over a limited range
(Fig.~\ref{fig-Pk}). Further, the log-normal term is relevant at
length scales $k_2$ determined by $\ln k_2\sim \sqrt{\ln
  n}$. Similarly there is a series of length scales $k_b$ that are
specified by $\ln k_b\sim (\ln n)^{(b-1)/b}$. For $k\gg k_b$, the
$b^\text{th}$ term in the Taylor expansion of $\Phi(x)$ affects the
length distribution. Ultimately, at sufficiently large scales, the 
entire scaling function \eqref{Phi} characterizes the length
distribution (Fig.~\ref{fig-Phi}).

In appendix A, we derive the exact length distribution 
\begin{equation}
\label{Pk-exact}
P_k = \frac{2}{k(k+1)}\,,
\end{equation}
which is realized in the limit $n\to\infty$. This distribution is
properly normalized, $\sum_{k\geq 2} P_k=1$, and its power-law tail
$P_k\simeq 2k^{-2}$ agrees with the asymptotic behavior
\eqref{Pk-tail}.  Furthermore, results of numerical evaluation of the
recursion equation \eqref{Smn-eq} with the boundary condition
\eqref{Skmn-bc} are in excellent agreement with this theoretical
prediction (Fig.~\ref{fig-Pk}).  The exact length distribution
\eqref{Pk-exact} can be expressed a ratio of Gamma functions, the
discrete counterpart of a power-law. It can be derived by treating the
variables $m$ and $n$ as discrete, in contrast with the continuum
analysis leading to \eqref{Pk-tail}.

\begin{figure}[t]
\includegraphics[width=0.425\textwidth]{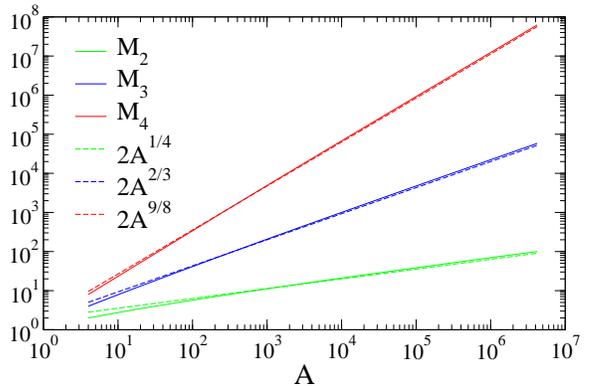}
\caption{The normalized moments $M_h$ versus square area $A$ for
  $h=2,3,4$. Also shown as a reference are the corresponding
  theoretical predictions.}
\label{fig-moments}
\end{figure}

Finally, we investigate the moments of the length distribution,
defined by $\langle k^h\rangle=\sum_{k\geq 2} k^h P_k$. It is
convenient to normalize these moments by the average length,
\begin{equation}
\label{Mh-def}
M_h=\frac{\langle k^h\rangle}{\langle k\rangle}\,,
\end{equation}
with $h>1$.  By using the definition $P_k=S_k/S$ and the second sum
rule in \eqref{norm}, we can express the normalized moments through
$S_k(n,n)$, the average number of frozen sticks with length $k$ in a
square of of size $n\times n$,
\begin{equation}
\label{Mh-sum}
M_h=n^{-2}\sum_{k\geq 2}  k^h S_k(n,n)\,.
\end{equation}
We now substitute \eqref{Snunuk} into this expression and convert the
sum over the discrete variable $k$ into an integral over the
continuous variable $x$ by using $k=e^{\nu x}$. With this
transformation of variables, the moments are given by
\begin{equation}
\label{Mh-int}
M_h\simeq \sqrt{\frac{\nu}{\pi}}\int_0^1 dx\,(1-x)^{-3/4} \, e^{\nu\phi(x)}\,,
\end{equation}
with $\phi(x)=\Phi(x)+(h+1)x$. The exponential term dominates the
integral  in the limit $\nu\to \infty$. The function $\phi(x)$ is
maximal at the saddle point, $x_*=1-h^{-2}$, and from the quadratic
behavior \hbox{$\phi(x)=\phi(x_*)+\tfrac{1}{2}\phi''(x_*)(x-x_*)^2$}, we
deduce the leading asymptotic behavior (Fig.~\ref{fig-moments})
\begin{equation}
\label{mu}
M_h\simeq 2\,A^{\mu(h)}\quad {\rm with}\quad \mu(h) =\frac{(h-1)^2}{2h}\,.
\end{equation}
Results of numerical evaluation of low-order moments are in
excellent agreement with this theoretical prediction. 

\begin{figure}[t]
\includegraphics[width=0.4\textwidth]{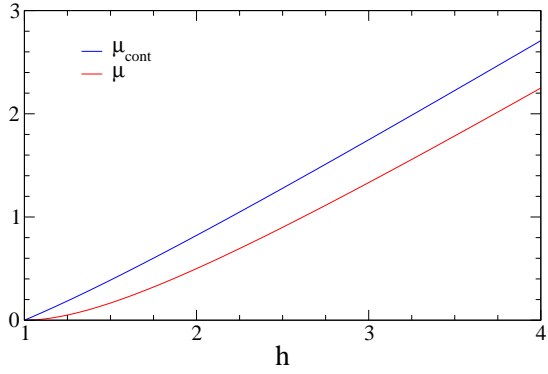}
\caption{The scaling exponents $\mu$ and $\mu_{\rm cont}$ versus the
  moment index $h$.}
\label{fig-cont}
\end{figure}

The scaling exponent $\mu$ is a nonlinear function of the index $h$,
and therefore, the scaling behavior of the first moment does not
characterize high-order moments. Hence, the moments exhibit
multi-scaling asymptotic behavior. Qualitatively similar behavior is
found for the continuous version of the fragmentation process
\eqref{process}. To compare the two cases, we note that the aspect
ratio of a frozen rectangle $r$ equals its length, $r=k$. For the
continuous version, the normalized moments of the aspect ratio,
$M_h=\langle r^h\rangle/\langle r\rangle$, also exhibit multi-scaling
asymptotic behavior $M_h\sim (A/\langle A\rangle)^{\mu_\text{cont}}$,
where $\langle A\rangle$ is the average rectangle area and
\begin{equation}
\label{mu-cont}
\mu_\text{cont}=\sqrt{h^2+1}-\sqrt{2}
\end{equation}
is the nonlinear scaling exponent \cite{kb,bk}. Surprisingly, the two
spectrums of exponents are different, $\mu\neq \mu_\text{cont}$
although both become linear at high orders, \hbox{$\mu\simeq
\mu_\text{cont}\simeq h$} as \hbox{$h\to\infty$}. Figure
\ref{fig-cont} shows that multi-scaling is more pronounced in the
discrete case.

\section{Asymmetric Fragmentation}
\label{sec:asym}

We now generalize the fragmentation process \eqref{process} and
consider the case where the probabilities of horizontal and vertical
cuts may differ \cite{tivp}. The asymmetric fragmentation process can
be represented schematically as
\begin{equation}
\label{asymmetric}
(m,n)\to 
\begin{cases}
(i,n)+(m-i,n)&   \text{prob.}~~(1-\alpha)/2, \\
(m,j)+(m,n-j)&   \text{prob.}~~(1+\alpha)/2.
\end{cases}
\end{equation}
The parameter $\alpha$ controls the degree of asymmetry, and without
loss of generality, we assume $0\leq \alpha\leq 1$.  The fragmentation
process \eqref{asymmetric} reduces to \eqref{process} when there is no
asymmetry, $\alpha=0$, and it becomes one-dimensional, when the
asymmetry parameter is maximal, $\alpha=1$.

In the completely asymmetric case, the jammed state contains $n$
identical sticks of length $n$.  Hence, the number of sticks is not
proportional to the area, and also, there is no logarithmic dependence
on system size, unlike \eqref{S}. Below, we show that the
logarithmic dependence on system size disappears when the asymmetry
parameter exceeds the critical value
\begin{equation}
\label{alphac}
\alpha_c=\frac{1}{\sqrt{2}}\,.
\end{equation}

The average number of frozen sticks  obeys 
\begin{equation}
\label{Smn-eq-a}
S(m,n)=\frac{1-\alpha}{m-1}\sum_{i=1}^{m-1} S(i,n) 
+ \frac{1+\alpha}{n-1}\sum_{j=1}^{n-1}S(m,j), 
\end{equation}
subject to the boundary condition \eqref{Smn-bc}. This recurrence
reduces to \eqref{Smn-eq} when the asymmetry parameter vanishes.

Once again, we employ the continuum approach.  In terms of the
logarithmic variables \eqref{munu}, the quantity \hbox{$S\equiv S(\mu,\nu)$} 
satisfies the partial differential equation
\begin{equation}
\label{Smunu-eq-a}
\partial_\mu \partial_\nu S = S+\alpha(\partial_\mu S-  \partial_\nu S)\,.
\end{equation}
We now repeat the steps leading to \eqref{Spq} and find that the 
double Laplace transform, defined in \eqref{Spq-def}, is given by 
\begin{equation}
\label{Spq-a}
\widehat{S}(p,q) = \frac{1+\alpha(p^{-1}-q^{-1})}{pq+\alpha(q-p)-1}\,.
\end{equation}
Since the governing equation \eqref{Smunu-eq-a} is no longer symmetric
in the variables $\mu$ and $\nu$, the Laplace transform \eqref{Spq-a}
is not symmetric when $\alpha\neq 0$.  We thus reiterate that results are
quoted only for squares.

\begin{figure}[t]
\includegraphics[width=0.425\textwidth]{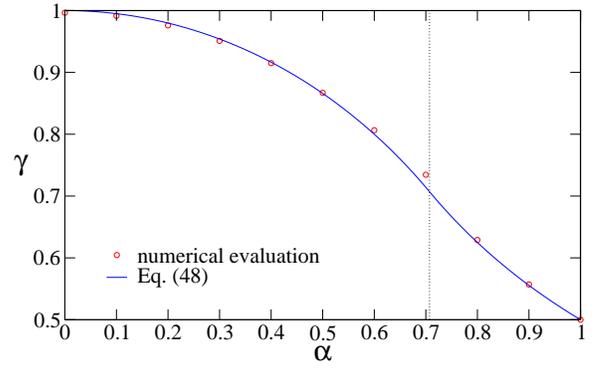}
\caption{The exponent $\gamma$ given by \eqref{gamma} versus the
  asymmetry parameter $\alpha$. Also shown are results of numerical
  evaluation of the recursion equations with $A=10^8$. To estimate the
  power-law exponent, we took into account the logarithmic correction
  in the weakly asymmetric phase, according to \eqref{U}. The critical
  point \eqref{alphac} is indicated by the dashed vertical line.}
\label{fig-gamma}
\end{figure}

To invert the Laplace transform \eqref{Spq-a}, we split the numerator
$1+\alpha(p^{-1}-q^{-1})$ into $q-$dependent and $p-$dependent terms:
$1-\alpha\,q^{-1}$ and $\alpha\,p^{-1}$.  Due to asymmetry, these two
are no longer equivalent, and the average number of sticks in the
jammed state $S\equiv S(\nu, \nu)$ is given by
\begin{equation}
\label{SIJ}
S = I+ J\,.
\end{equation}
The quantity $I\equiv I(\nu,\nu)$ is obtained by inverting
\hbox{$\big(1-\alpha\,q^{-1}\big)/[pq+\alpha(q-p)-1]$} first with
respect to the conjugate variable $p$ and then with respect to the
conjugate variable $q$.  Similarly, the quantity $J\equiv J(\nu,\nu)$
is obtained by inverting \hbox{$\alpha\,p^{-1}/[pq+\alpha(q-p)-1]$}
first with respect to $q$ and then, with respect to $p$.

To compute the first term in \eqref{SIJ}, we follow the calculations
in the symmetric case, and find that $I$ is an integral of the form
\eqref{I-def}, specified by $F(q)=q^{-1}$ and
\begin{equation}
\label{f-a}
f(q) = q-\alpha +\frac{\beta^2}{q-\alpha}\,, \qquad \beta=\sqrt{1-\alpha^2}\,.
\end{equation}
The saddle point is simply $q_*=\beta+\alpha$, and from the general
formula \eqref{I-gen} we obtain
\begin{equation}
\label{I}
I\simeq \frac{\beta}{\beta + \alpha}\,\,\frac{e^{2\beta\nu}}{\sqrt{4\pi\beta\nu}}\,.
\end{equation}
This asymptotic behavior resembles \eqref{S} in that the leading exponential
behavior is suppressed by a logarithmic term.

The second quantity in \eqref{SIJ} is analogous in form to
\eqref{I-def}: it is given by a integral over $p$, rather than $q$,
\begin{equation}
\label{J-def}
J=\int_{-\ii \infty}^{\ii \infty} \frac{dp}{2\pi \ii}\,G(p)\,e^{\nu g(p)}\,,
\end{equation}
with the functions
\begin{equation}
\label{Gg}
G(p)=\frac{\alpha}{p(p+\alpha)},\qquad g(p)= p+\alpha +\frac{\beta^2}{p+\alpha}\,.
\end{equation}
However, the saddle point of the function $g(p)$ is different, $p_* =
\beta -\alpha$. To evaluate the integral $J$, we simply replace
$F(q_*)$, $f(q_*)$, and $f''(q_*)$ in \eqref{I-gen} with $G(p_*)$,
$g(p_*)$, and $g''(p_*)$ respectively, to find
\begin{equation}
\label{J}
J\simeq \frac{\alpha}{\beta-\alpha}\,\,\frac{e^{2\beta\nu}}{\sqrt{4\pi\beta\nu}}\,.
\end{equation}
Therefore, the two terms in the sum \eqref{SIJ} are proportional to
each other, and by adding \eqref{I} and \eqref{J}, we arrive at
\begin{equation}
\label{S-sub}
S\simeq C\,  \,\frac{e^{2\beta\nu}}{\sqrt{4\pi\beta\nu}}\,, \qquad 
C = \frac{1}{2(\alpha_c^2-\alpha^2)}\,,
\end{equation}
with the critical point $\alpha_c$ given in \eqref{alphac}. The
constant $C$ diverges as $\alpha \uparrow \alpha_c$, thereby
indicating that the result \eqref{S-sub} is valid only when the
asymmetry is sufficiently weak, $\alpha<\alpha_c$. Indeed, the
integrand in \eqref{J-def} has two simple poles: one at $p=-\alpha$
and another at $p=0$. The first pole is located to the left of the
saddle point $p_*$, irrespective of $\alpha$. However, the second pole
is located to the left of the saddle point only when
$\alpha<\alpha_c$, and consequently, Eq.~\eqref{J} holds only in this
regime.

\begin{figure}[t]
\includegraphics[width=0.425\textwidth]{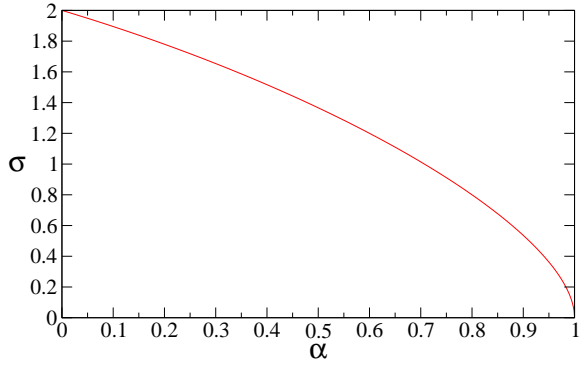}
\caption{The power-law exponent $\sigma$ versus the asymmetry 
  parameter $\alpha$.}
\label{fig-sigma}
\end{figure}

To evaluate the integral $J$ when $\alpha>\alpha_c$, we deform the
integration contour so that it consists of a line parallel to the
imaginary axis which passes through the saddle point $p_*$ and a small
circle enclosing the origin, $p=0$. The residue at the origin gives
the dominant contribution, $J\simeq e^{\nu/\alpha}$, which is valid
when the asymmetry is sufficiently strong, \hbox{$\alpha
  >\alpha_c$}. In this regime, the quantity $I$ in \eqref{I} is
negligible, and consequently, $S(\nu,\nu)\simeq e^{\nu/\alpha}$.

In summary, the number of sticks in the jammed state state grows
algebraically with  area $A$ 
\begin{equation}
\label{gamma}
S\simeq U\,A^{\gamma},\quad   {\rm with}\quad
\gamma = 
\begin{cases}
\sqrt{1-\alpha^2}                 & \alpha \leq \alpha_c\,,\\
1/(2\alpha)                         & \alpha \geq \alpha_c \,.
\end{cases}
\end{equation}
As long as there is some asymmetry, the growth is sub-linear:
$\gamma<1$ when $\alpha>0$.  The exponent $\gamma$ is continuous at
the critical point, but its first derivative is discontinuous at that
point. Furthermore, the exponent $\gamma$ is concave when $\alpha <
\alpha_c$ but it is convex when $\alpha>\alpha_c$. The critical value
is $\gamma_c=\alpha_c$, and the convergence toward the leading
asymptotic behavior is slower near the critical point
(Fig.~\ref{fig-gamma}).

The prefactor $U\equiv U(A)$ in \eqref{gamma} depends logarithmically
on area in the weakly asymmetric phase, and it is given by
\begin{equation}
\label{U}
U=
\begin{cases}
C/\sqrt{2\pi\beta\ln A}       \quad & \alpha<\alpha_c,\\
\sqrt{e/(4\pi)}               \quad & \alpha=\alpha_c,\\
1                             \quad & \alpha>\alpha_c\,.
\end{cases}
\end{equation}
The constant $C$ is quoted in \eqref{S-sub}, and the behavior at the
critical point is derived in Appendix B.  Hence, there is logarithmic
correction in the weakly asymmetric phase, but the logarithmic
correction disappears in the strongly asymmetric phase.  The critical
prefactor $U_c=\sqrt{e/(4\pi)}$ obeys $0<U_c<1$, and we note the
``double discontinuity'': the critical value $U_c$ does not match
either of the limiting behaviors $U\to 0$ as $\alpha\uparrow \alpha_c$
or $U\to 1$ as $\alpha\downarrow \alpha_c$.

\begin{figure}[t]
\includegraphics[width=0.425\textwidth]{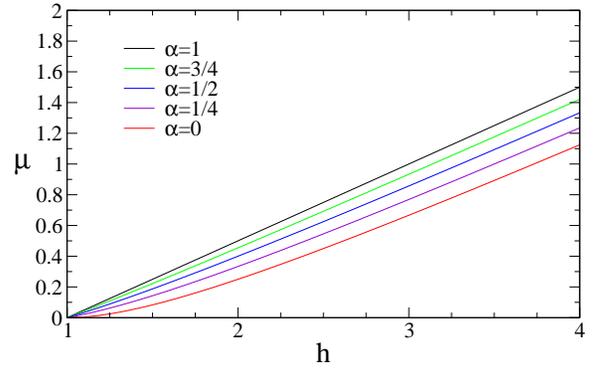}
\caption{The scaling exponent $\mu$ versus the moment index $h$ for
  various values of the asymmetry parameter $\alpha$.}
\label{fig-mua}
\end{figure}

The length distribution decays algebraically 
\begin{equation}
\label{sigma}
P_k\simeq V\,k^{-\sigma},\quad{\rm with}\quad \sigma=1+\beta-\alpha\,,
\end{equation}
for $k\gg 1$. This behavior is derived in Appendix C.  The power-law
tail \eqref{sigma} generally holds for infinitely large
systems. However, for finite systems, this behavior holds in the range
$\ln k\ll \sqrt{\ln n}$, as discussed above.  Generally, the exponent
$\sigma$ decreases monotonically as $\alpha$ increases, and it
vanishes in the completely asymmetric case
(Fig.~\ref{fig-sigma}). Therefore, as the asymmetry parameter becomes
smaller, the tail of the length distribution decays more sharply
(see also Ref.~\cite{tivp}).

Interestingly, the exponent $\sigma$ which characterizes the tail of
the length distribution has the same form \eqref{sigma} in the weakly
asymmetric phase, $\alpha\leq \alpha_c$ and the strongly asymmetric
phase $\alpha\geq \alpha_c$. Yet, the prefactor $V\equiv V(A)$ in
\eqref{sigma} depends algebraically on area in the strongly asymmetric
phase, and it is given by
\begin{equation}
\label{V}
V=
\begin{cases}
2\beta(\alpha_c^2-\alpha^2)                     \quad  & \alpha<\alpha_c\,,\\
2^{1/4}/\sqrt{(e \ln A)}                         \quad   & \alpha=\alpha_c\,,\\
\sqrt{\beta/(2\pi \ln A)}\, A^{\beta-1/(2\alpha)}   \quad   &\alpha>\alpha_c\,.
\end{cases}
\end{equation}
Therefore, the length distribution depends on system size in the
strongly asymmetric phase, but it is independent of system size in the
weakly asymmetric phase. 

From the length distribution $P_k$, it is also possible to evaluate the
moments $M_h$ defined in \eqref{Mh-def}.  We find that the moments
grow algebraically with the area as in \eqref{mu}, $M_h\sim
A^\mu$. For asymmetric fragmentation, the spectrum of scaling
exponents is given by
\begin{equation}
\label{mu-a}
\mu = \frac{\beta^2}{2(h+\alpha)}+\frac{h+\alpha}{2} - 1\,.
\end{equation}
This spectrum reduces to \eqref{mu} when fragmentation is symmetric,
but in contrast, the scaling exponents are linear,
$\mu=\tfrac{h-1}{2}$, when fragmentation is completely asymmetric. 
Figure \ref{fig-mua} demonstrates how multi-scaling becomes
less pronounced as the asymmetric nature of the fragmentation process
becomes stronger.

\begin{figure}[t]
\includegraphics[width=0.45\textwidth]{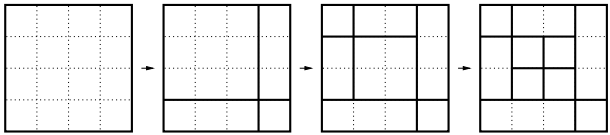}
\caption{Illustration of the fragmentation process
  \eqref{deterministic}.  Initially, the system consists of a single
  rectangle. Through a series of random fragmentation events, the
  system arrives at a jammed state where all rectangles are sticks
  with minimal horizontal or vertical size.}
\label{fig-ill2}
\end{figure}

\section{Deterministic Fragmentation}

The fragmentation process \eqref{process} incorporates two stochastic
elements as both the fragmentation point and the fragmentation
direction are selected at random.  The latter element can be
eliminated by generating four rectangles, rather than two, in each
fragmentation event (Fig.~\ref{fig-ill2}).  A continuous version of
this planar fragmentation process was introduced in \cite{tv} and
analyzed in a number of subsequent studies
\cite{kb,rh,btv,btv97,bk97}.

\begin{figure}[t]
\includegraphics[width=0.35\textwidth,height=0.35\textwidth]{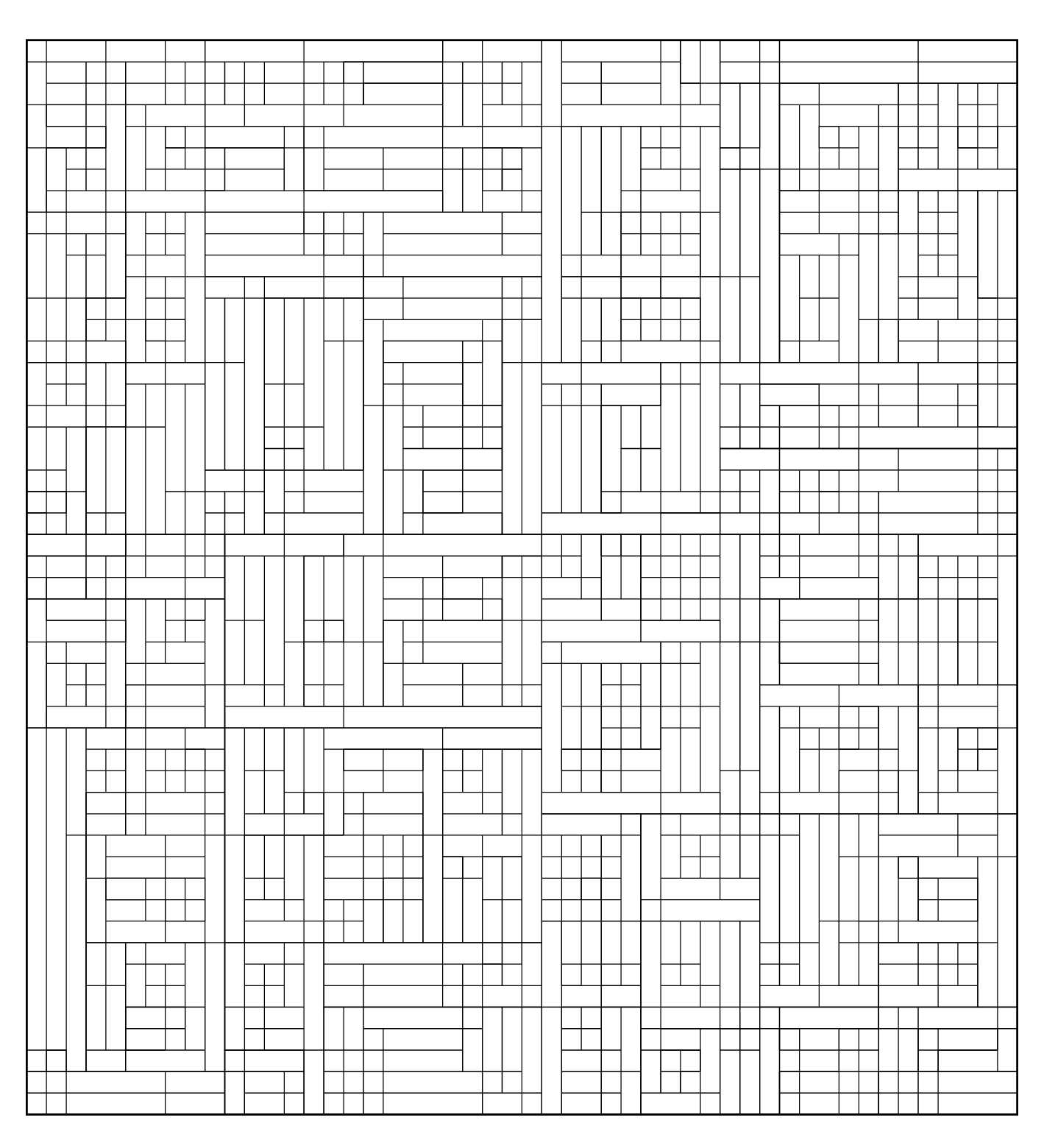}
\caption{A jammed configuration in a system of size $50\times 50$.}
\label{fig-jm}
\end{figure}

We now address this natural counterpart of the fragmentation process
\eqref{process} where first an internal grid point is selected at
random, and then, two simultaneous cuts are made, one in the
horizontal direction and one in the vertical direction.  As a result,
each fragmentation event generates four rectangles
(Fig.~\ref{fig-ill2})
\begin{equation}
\label{deterministic}
(m,n)\to (i,j)+(m-i,j)+(i,n-j)+(m-i,n-j)\,,
\end{equation}
with randomly selected $1\leq i\leq n-1$ and $1\leq j\leq m-1$.  In
contrast with \eqref{process}, once the grid point is selected, the
outcome is deterministic.  Again, rectangles with \hbox{$m>1$} and
\hbox{$n>1$} are active and otherwise, rectangles with \hbox{$m=1$} or
$n=1$ are frozen.  Starting with a single $m\times n$ rectangle, the
system eventually reaches a jammed state where all rectangles are
sticks, including minimal $1\times 1$ rectangles (Fig.~\ref{fig-jm}).

The average number of frozen sticks in the jammed state, $S(m,n)$, 
satisfies the recursion equation 
\begin{equation}
\label{Smn-eq-d}
S(m,n)=\frac{4}{(m-1)(n-1)}\sum_{i=1}^{m-1}\sum_{j=1}^{n-1} S(i,j)\,,
\end{equation}
subject to the boundary conditions \eqref{Smn-bc}. For small
rectangles, the recursion equation gives $S(2,2)=4$, $S(3,3)=7$,
$S(4,4)=\frac{32}{3}$, and so on.  In contrast with the fragmentation
process \eqref{process} where the average number of sticks for narrow
but long rectangles diverge logarithmically, these quantities are now
finite, and for example, $S(2,n)=4$ and $S(3,n)=10-\tfrac{6}{n-1}$.  In
general, we find the limiting values
\begin{equation}
\label{Smn-large-d}
\lim_{n\to\infty} S(m,n) = \frac{m(m+1)(m+2)}{6}\,,
\end{equation}
from the recursion equation \eqref{Smn-eq-d}.

\begin{figure}[t]
\includegraphics[width=0.425\textwidth]{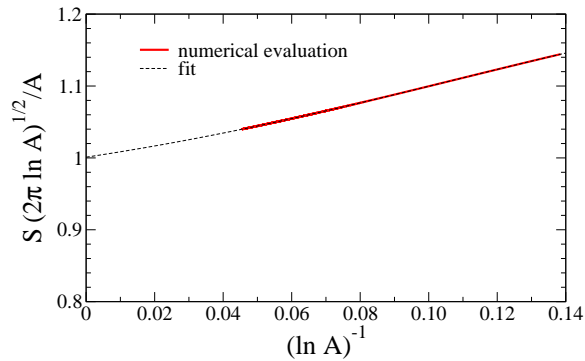}
\caption{The quantity $S\sqrt{2\pi \ln A}/A$ versus $(\ln
  A)^{-1}$. The dashed line shows results of a fourth-order polynomial
  fit to the data. Results of numerical evaluation of the recursion
  equation \eqref{Smn-eq-d} agrees with the theoretical prediction for
  the leading asymptotic behavior to within 0.1\%.}
\label{fig-S-d}
\end{figure}

For large rectangles, the recursion equation \eqref{Smn-eq-d} turns
into the partial differential equation
\begin{equation}
\label{Smn-eq1-d}
\partial_m \partial_n [m n S(m,n)] = 4 S(m,n)\,.
\end{equation}
Using the logarithmic variables \eqref{munu}, we transform this
equation into a partial differential equation with constant
coefficients, \hbox{$\partial_\mu \partial_\nu S + \partial_\mu S +
  \partial_\nu S = 3S$}.  By repeating the steps leading to
\eqref{Spq}, we obtain the Laplace transform
\begin{equation}
\label{Spq-d}
\widehat{S}(p,q) = \frac{1}{pq}\,\frac{(p+1)(q+1)-1}{(p+1)(q+1)-4}\,.
\end{equation}
Next, we rewrite this expression as a sum of two terms:
$(2^{-1}+q^{-1})/(pq+p+q+3)$ and $(2^{-1}+p^{-1})/(pq+p+q+3)$. For
squares, these two terms are equivalent and it suffices to perform the
inverse Laplace transform of the first term with respect to $p$ and  
then, with respect to $q$.  We thus obtain $S(\nu,\nu)=I$ where $I$ is
given by the general integral \eqref{I-gen}. The integrand is
specified by the functions
\begin{equation}
\label{Ff-d}
F(q) = \frac{q+2}{q(q+1)}\quad{\rm and}\quad f(q) = q-1+\frac{4}{q+1}\,.
\end{equation}
From the condition $f'(q_*) = 0$ we notice that the saddle point
remains the same, $q_*=1$, and by using \eqref{I-gen}, we find the
leading asymptotic behavior of the number of frozen sticks in the
jammed state (Fig.~\ref{fig-S-d})
\begin{equation}
\label{S-d}
S\simeq \frac{3 A}{\sqrt{4\pi \ln A}}\,.
\end{equation}
In comparison with \eqref{S}, the average number of frozen sticks is
now $3/\sqrt{2}\approx 2.12132$ times larger.  Results of numerical
evaluation of the recursion equation \eqref{Smn-eq-d} are in excellent
agreement with the theoretical prediction \eqref{S-d}.  As was the
case for stochastic fragmentation, the average number of jammed
rectangles \eqref{S-d} extends to all rectangles with a finite aspect
ratio in the large-area limit.

For completeness, we quote the exact distribution of stick length 
\begin{equation}
\label{Pk-exact-d}
P_k = \frac{4}{3k(k+1)}\,,
\end{equation}
for $k>1$ and $P_1=\frac{1}{3}$. This form, which is realized in the
limit $n\to\infty$, can be obtained using the method outlined in
Appendix A. Sticks with $k\geq 2$ are doubly degenerate compared with
minimal $1\times 1$ rectangles and hence, the quantity $P_1$ is
suppressed by a factor 2. The length distribution has a power-law
tail, $P_k\simeq \tfrac{4}{3}k^{-2}$, which can be established using 
continuum analysis used to obtain \eqref{Pk-tail}.  For a finite
system, the power-law tail holds when $1\ll k\ll \sqrt{\ln n}$, while
at larger length scales the distribution is strongly suppressed by a
log-normal term.

\begin{figure}[b]
\includegraphics[width=0.2\textwidth]{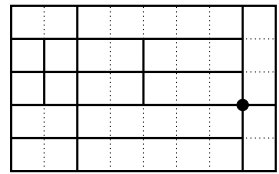}
\caption{Illustration of the fragmentation process \eqref{deterministic}. 
The first fragmentation event can always be uniquely identified.}
\label{fig-cut4}
\end{figure}

The planar fragmentation processes considered in this investigation
generate special tilings of two-dimensional domains. Indeed, in the
jammed configuration, sticks of unit width and variable length cover
the original rectangle (Figures \ref{fig-jam} \& \ref{fig-jm}). The
jammed configurations differ from those in the heavily studied dimer
tiling \cite{pwk,ft,mef,ehl,rjb,cep,ckp} in two respects. First, the
lengths of the sticks do vary dramatically
\cite{hl,mj,fw,gjl}. Second, whereas in equilibrium problems jammed
configurations are given equal weights, fragmentation is a dynamical
process, and the different tiling configurations are generally
realized with different probabilities. In this sense, the jammed
configurations considered here can be viewed as nonequilibrium
tilings.

The central quantity in tiling problem is the total number of jammed
configurations which typically grows exponentially with area.  For the
deterministic process \eqref{deterministic}, it is straightforward to
show that $T(m,n)$, the total number of jammed configurations for a
rectangle of size $m\times n$, satisfies the recursion equation
\begin{equation}
\label{T-rec}
T(m,n)\!= \!\!\!\!\!\sum_{1\leq i\leq m-1\atop 1\leq j\leq n-1}\!\!\!\!\!
T(i,j)T(m-i,j)T(i,n-j)T(m-i,n-j).
\end{equation}
This recursion applies for all $m\geq 2$ and $n\geq 2$, and it is
subject to the boundary conditions $T(m,1)=1$ for all $m\geq 1$ and
$T(1,n)=1$ for all $n\geq 1$. As illustrated in Fig.~\ref{fig-cut4},
for any jammed configuration, the first fragmentation event can be
uniquely identified.  This first fragmentation event divides the
original rectangle to four smaller and independent rectangles, thereby
leading to the recursion \eqref{T-rec}. The same does not hold true
for the fragmentation process \eqref{process} and for this reason, it
is not possible to write closed recursion equations for the
corresponding of number of jammed states.

\begin{figure}[t]
\includegraphics[width=0.425\textwidth]{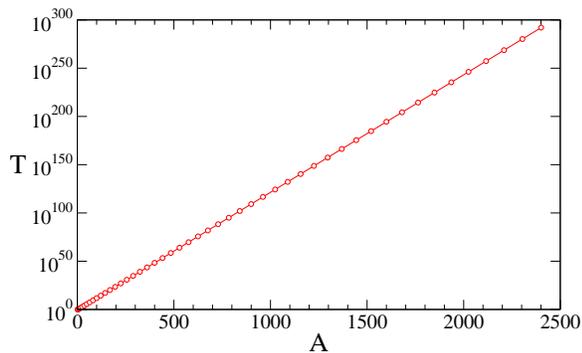}
\caption{The total number of jammed configurations for squares, $T=T(n,n)$,
 versus the area $A=n^2$.}
\label{fig-t}
\end{figure}

A single iteration of the recursion \eqref{T-rec} yields the number of
jammed configurations for ladders, $T(2,n) = n-1$, and a second
iteration yields \hbox{$T(3,n) = \tfrac{1}{3}(n-2)(n^2-4n+15)$}.  The
exact expression for $T(4,n)$ is a seventh-order polynomial, and
$T(m,n)$ quickly become unwieldy when $m$ increases.

Table I lists the number of jammed configurations for squares with
$n\leq 7$.  Numerical iteration of the recursion \eqref{T-rec} shows
that the number of jammed configurations grows exponentially with 
area (see Fig.~\ref{fig-t})
\begin{equation}
\label{T}
T\sim e^{\lambda A}.
\end{equation}
For squares, we obtain $\lambda=0.2805$ by fitting the quantity
$T(n,n)$ to an exponential. We also mention another numerical
observation: the numeric prefactor in \eqref{T} has finite upper and
lower bounds.

\begin{table}[t]
%\vspace*{0.66cm}
\begin{tabular}{|c|c|c|c|c|c|c|c|c|}
\hline
$n$         &   1 & 2 & 3 & 4 & 5 & 6 & 7 & 8 \\
\hline
$T(n,n)$ & 1 & 1 & 4 & 33 & 436 & 9,524 & 354,224 & 23,097,969\\
\hline
\end{tabular}
\caption{The number of jammed states for small squares.} 
\label{Tab:jammed}
\end{table}

\section{Discussion}

In summary, we studied planar fragmentation, which can be viewed as
dual to planar aggregation \cite{bbk}.  We obtained analytically
several properties of the jammed state including the average number of
rectangles, and the length distribution of rectangles. In general,
statistical properties become independent of the aspect ratio in the
large-area limit. Moreover, the length distribution of rectangles in
the jammed state has a power-law tail, and the moments of this
distribution exhibit multiscaling.

We also found a phase transition when the fragmentation process is
asymmetric.  Generally, the average number of jammed rectangles grows
sub-linearly with system size, and the exponent characterizing this
growth varies continuously with the asymmetry parameter. This exponent
is concave in the weakly asymmetric phase and convex in the strongly
asymmetric phase. In addition, The length distribution is independent
of system size in the weakly asymmetric phase, but it does depend
on system size in the strongly asymmetric phase.

Our theoretical analysis relies on recursion equations that describe
the final state of the system. Since each fragmentation event involves
a single rectangle, the recursion equations are linear. For large
systems, we employed the continuum approach and then applied the
Laplace transform to obtain exact results for the leading asymptotic
behavior.  Numerical evaluation of the recursion equations
support the theoretical predictions.

The recursion equations bypass the evolution toward the jammed state
and hence, directly yield statistics of the final configuration. The
fragmentation rate may be an arbitrary function of the area, yet,
as long as the fragmentation point is selected at random, the
recursion equation \eqref{Smn-eq} holds. For specific fragmentation
rates, it is natural to study the evolution towards the jammed state,
including in particular the average jamming time, and the distribution
of jamming times.

Our analysis yields statistics of {\em single} fragments in the jammed
states such as the average length and the length distribution.
Missing from our analysis, however, are statistics of {\em multiple}
fragments such as correlations between the orientations of neighboring
sticks. Both sets of statistics are relevant for characterizing the
geometrical structure of planar fragmentation patterns found in
martensitic transformations \cite{tivp}, breakage of brittle objects
\cite{vv,dp}, cracking of soils \cite{bpc}, and drying of suspensions
\cite{nm}.

The behavior in higher dimensions can be studied as well.  In the
three-dimensional generalization of \eqref{process}, the jammed state
consists of rectangular plates, that is, boxes with unit width.  In
this case, we find that the average number $F$ of frozen boxes grows
as
\begin{equation}
F\simeq \left(\frac{\sqrt{3}}{2}\right)^3\frac{V}{\pi \ln V}\,,
\end{equation}
with $V$ the volume of the original box. The area distribution of
jammed plates represents an interesting challenge.

The deterministic process \eqref{deterministic} can be also
generalized to $d$ dimensions. Here, each fragmentation event
generates $2^d$ boxes. The jammed state consists of frozen boxes, each
of which has at least one minimal side.  The number of frozen boxes
$F$ grows as
\begin{equation}
\label{F-d}
F\simeq \frac{2^d-1}{\sqrt{d}}\, 
\frac{V}{\left(\tfrac{4\pi}{d}\,\ln V\right)^{(d-1)/2}}\,.
\end{equation}
This result, which generalizes \eqref{S-d}, is derived in Appendix D.
A frozen box is characterized by $d-1$ nontrivial lengths and it is an
interesting challenge to characterize the distribution of these
lengths.

\bigskip
We thank Jean-Marc Luck for making helpful comments.  This work was
supported in part by the US Department of Energy through the Los
Alamos National Laboratory. Los Alamos National Laboratory is operated
by Triad National Security, LLC, for the National Nuclear Security
Administration of U.S. Department of Energy (Contract
No. 89233218CNA000001).

\appendix

\section{Length Distribution}

To obtain the length distribution for an infinite system, we treat $m$
and $n$ as {\rm discrete} variables. To this end, we introduce the
generating function
\begin{equation}
\label{Sxy}
\mathcal{S}(x,y)=\sum_{m\geq 1}\sum_{n\geq 1} S(m,n)\, x^{m-1} y^{n-1}\,.
\end{equation}
The generating function $\mathcal{S}\equiv\mathcal{S}(x,y)$ satisfies the partial
differential equation
\begin{equation}
\label{Sk:PDE}
\partial_x\partial_y \mathcal{S} =
(1-x)^{-1}\,\partial_y\mathcal{S}+
(1-y)^{-1}\,\partial_x\mathcal{S}\,.
\end{equation}
To obtain this equation, we multiply the recursion equation
\eqref{Smn-eq}, that governs the averages $S(m,n)$, by
\hbox{$(m-1)(n-1)x^{m-2} y^{n-2}$} and sum over $m\geq 2$ and $n\geq
2$. Furthermore, equation \eqref{Sk:PDE} is subject to the boundary conditions
$\mathcal{S}(x,0) = x/(1-x)$ and $\mathcal{S}(0,y) = y/(1-y)$.

We now introduce the variables
\begin{equation}
\label{xi-eta}
\xi = -\ln(1-x), \quad \eta =  -\ln(1-y)\,.
\end{equation}
This transformation turns \eqref{Sk:PDE} into a partial differential
equation with constant coefficients
\begin{equation}
\label{Sk:eq}
\partial_\xi \partial_\eta \mathcal{S} =
\partial_\xi\mathcal{S} + \partial_\eta\mathcal{S}\,,
\end{equation}
while the boundary conditions become $\mathcal{S}(\xi,0) = e^\xi-1$
and $\mathcal{S}(0,\eta) = e^\eta-1$.  Next, we introduce the double
Laplace transform
\begin{eqnarray}
\label{Lap-S:def}
\widehat{\mathcal{S}}(p,q) &=& \int_0^\infty d\xi\,e^{-p\xi} \int_0^\infty d\eta\,e^{-q\eta}\, \mathcal{S}(\xi,\eta)\,.
\end{eqnarray}
Using the governing equation \eqref{Sk:eq}, we obtain 
\begin{eqnarray}
\label{Spq1}
\widehat{S}(p,q)=\frac{p^{-1}+q^{-1}}{pq-p-q}.
\end{eqnarray}
We now invert the Laplace transform with respect to one of the conjugate
variables, to obtain the sum 
\begin{eqnarray*}
\mathcal{S}(\xi,\eta) 
&=& \int_{-\ii \infty}^{\ii \infty} \frac{dq}{2\pi \ii}\,\,\frac{1}{q(q-1)}\,e^{\eta q+\xi q/(q-1)}\\
&+& \int_{-\ii \infty}^{\ii \infty} \frac{dp}{2\pi \ii}\,\,\frac{1}{p(p-1)}\,e^{\xi p+\eta p/(p-1)}\,.
\end{eqnarray*}
Thus far, our analysis is exact and in particular, it applies to all
$m$ and $n$. We now restrict our attention to squares, $\xi=\eta$ and
further, we focus on the leading asymptotic behavior for large systems
which is captured by the leading behavior when $\eta\to\infty$.  By
performing the inverse Laplace transform over the second conjugate
variable, we obtain
\begin{equation}
\label{See}
S(\eta,\eta)\simeq \frac{e^{4\eta}}{\sqrt{4\pi\eta}}.
\end{equation}

Next, we analyze the length distribution $S_k(m,n)$. Its corresponding
generating function satisfies Eq.\eqref{Sk:eq}, subject to the
boundary conditions $\mathcal{S}_k(\xi,0)=(1-e^{-\xi})^{k-1}$
$S_k(0,\eta)=(1-e^{-\eta})^{k-1}$. These two boundary conditions
follow from $\mathcal{S}_k(x,0)=x^{k-1}$ and
$\mathcal{S}_k(0,y)=y^{k-1}$.  By repeating the steps leading to
\eqref{Spq1}, we obtain 
\begin{eqnarray*}
\label{Lap-Sk:def}
\widehat{\mathcal{S}}_k(p,q) 
=\frac{\Gamma(k)}{pq-p-q}\left[\frac{(q-1)\Gamma(q)}{\Gamma(k+q)} + \frac{(p-1)\Gamma(p)}{\Gamma(k+p)} \right].
\end{eqnarray*}
We can verify that $\mathcal{S}(p,q)=\sum_{k\geq
  2}\mathcal{S}_k(p,q)$, whereas the corresponding quantity in
\eqref{Skpq}, which is obtained by treating the variables $m$ and $n$
as continuous, violates this normalization.  The leading asymptotic
behavior in the limit $\eta\to \infty$ is given by 
\begin{equation}
\label{skee}
\mathcal{S}_k(\eta,\eta)\simeq \frac{2}{k(k+1)}\frac{e^{4\eta}}{\sqrt{4\pi\eta}}\,.
\end{equation}
The average number of sticks does not depend on aspect ratio and thus,
we assume $S_k(m,n)\simeq P_kS(m,n)$ at large sizes.  Then, according
to the definition \eqref{Sxy} and the leading asymptotic behavior
\eqref{See} we have \hbox{$\mathcal{S}_k(\eta,\eta)\simeq
  P_ke^{4\eta}/\sqrt{4\pi\eta}$}. By comparing this expression with
\eqref{skee}, we deduce the length distribution $P_k$ in
\eqref{Pk-exact}.

\section{Critical Behavior}

Here, we derive the critical behavior for the asymmetric fragmentation
process \eqref{asymmetric}.  The average number of jammed sticks is
generally given by the sum \eqref{SIJ}. First, we compute the integral
$J$ defined in \eqref{J-def}.  At the critical point,
$\alpha=\alpha_c$, the factor $p^{-1}$ in the integrand \eqref{Gg}
becomes large near the saddle point, and thus, we incorporate this
term into the exponential
\begin{equation}
J =\int_{-\ii \infty}^{+\ii \infty} 
\frac{dp}{2\pi \ii}\,\frac{\alpha}{p+\alpha}\,e^{\nu g(p)-\ln p}\,.
\end{equation}
The saddle point $p_*$ vanishes algebraically with $\nu$,  
\begin{equation}
\label{window-p}
p_*\simeq 2^{-3/4}\nu^{-1/2},
\end{equation}
in the $\nu\to\infty$ limit.  To perform the integration, we again
chose an integration contour parallel to the imaginary axis which
passes through $p_*$ and compute
\begin{equation}
\label{Jc}
J\simeq \sqrt{\frac{e}{4\pi}}\,e^{\sqrt{2}\,\nu}\,. 
\end{equation}
Next we evaluate Eq.~\eqref{I} at the critical point. The value 
$I\simeq 2^{-3/4} (4\pi\nu)^{-1/2} e^{\sqrt{2}\,\nu}$ is negligible
compared with $J$, and thus, $S\simeq J$ with $J$ given by \eqref{Jc}.

For infinite systems, the critical behavior applies strictly at the
critical point.  For finite systems, the critical behavior
characterizes a small, yet finite, region near the critical point.
The size of this region, which often referred to as the ``scaling window,''
shrinks as the system size increases, as follows from 
the saddle point \eqref{window-p},
\begin{equation} 
|\alpha - \alpha_c| \sim \nu^{-1/2}  \,.
\end{equation}
The scaling window decays very slowly with system size, and hence, it
is relevant even for reasonably large system.

\section{Asymmetric fragmentation}

The average number of sticks with a given length satisfies the 
partial differential equation \eqref{Smn-eq} subject to the boundary
conditions \eqref{Skmunu-bc}.  The double Laplace transform, defined
by \eqref{Spq-def}, is given by
\begin{equation*}
\label{Skpq-a}
\widehat{S}_k(p,q)=\frac{p k^{-1-p}+q k^{-1-q}+\alpha(k^{-1-p} - k^{-1-q})}{pq+\alpha(q-p)-1}\,.
\end{equation*}
To perform the double inverse Laplace transform, we rewrite the
numerator as a sum of the $q$-dependent quantity $(q-\alpha)k^{-1-q}$
and the $p$-dependent quantity $(p+\alpha)k^{-1-p}$. The quantity
$S_k(p,q)$ is therefore a sum of two terms as in \eqref{SIJ}.

The first term in the sum \eqref{SIJ} is the integral $I$ defined in
\eqref{I-def} with $F(q)=1$ and
\begin{equation}
f(q)= q-\alpha +\frac{\beta^2}{q-\alpha}-(1+q)x\,.
\end{equation}
Here, we again used the notations \hbox{$\beta=\sqrt{1-\alpha^2}$} and
\hbox{$x=\ln k/\ln n$}.  The saddle point of the function $f(q)$ is
$q_*=\beta(1-x)^{-1/2}+\alpha$ and with \eqref{I-gen}, we arrive at
\begin{eqnarray}
I&=&\frac{\exp\left[\nu\left(2\beta\sqrt{1-x}-x-\alpha x\right)
\right]}{\sqrt{4\pi\nu(1-x)^{3/2}/\beta}}\nonumber\\
\label{Ik}
 &\simeq& \frac{\sqrt{\beta}}{\sqrt{4\pi\nu}}\,e^{2\beta\nu}\,k^{-(1+\beta+\alpha)}\,.
\end{eqnarray}
By evaluating the small-$x$ behavior of the general expression in the
first line, we obtained the large-$k$ behavior in the second line.

The second term in the sum \eqref{SIJ} is the integral $J$ defined in
\eqref{J-def} with $G(p)=1$ and
\begin{equation}
g(p)= p+\alpha +\frac{\beta^2}{p+\alpha}-(1+p)x\,.
\end{equation}
The saddle point of the function $g(p)$ is
\hbox{$p_*=\beta(1-x)^{-1/2}-\alpha$}. By using the analog of the
general expression \eqref{I-gen}, we obtain
\begin{eqnarray}
\label{Jk}
J&=&\frac{\exp\left[\nu\left(2\beta\sqrt{1-x}-x+\alpha x\right)\right]}
{\sqrt{4\pi\nu(1-x)^{3/2}/\beta}}\nonumber\\
 &\simeq& \sqrt{\frac{\beta}{4\pi\nu}}\,e^{2\beta\nu}\,k^{-(1+\beta-\alpha)}\,.
\end{eqnarray}
The large-$k$ behavior in the second line follows from the small-$x$
behavior in the first line.  By comparing the tails \eqref{Ik} and
\eqref{Jk}, we conclude that $I$ is negligible compared with $J$, and
therefore $S_k(\nu,\nu)\simeq J$ for sufficiently large $k$.  The
power-law tail \eqref{sigma} follows from
\hbox{$P_k=S_k(\nu,\nu)/S(\nu,\nu)$}.

\section{Arbitrary Dimensions}
\label{ap:F-d}

The deterministic process \eqref{deterministic} can be generalized to
$d$ dimensions where in each fragmentation event a box breaks into
$2^d$ boxes. This elementary event is repeated until a jammed state is
reached. The recursion equation for the number of frozen boxes is a
straightforward generalization of Eq.~\eqref{Smn-eq-d} and it includes
$d$ sums.

The multivariate Laplace Transform is given by a straightforward
generalization of \eqref{Spq-d}
\begin{equation}
\label{Lap:d}
\widehat{S}(q_1,q_2,\ldots,q_d) =
\frac{\prod_{\ell=1}^d (q_\ell+1)-1}{\prod_{\ell=1}^d (q_\ell+1)
  -2^d}\times \prod_{\ell=1}^d  \frac{1}{q_\ell}\,.
\end{equation}
The inverse Laplace transform is a $d-$fold integral, and we first
invert this multivariate transform with respect to the conjugate
variable $q_d$. Further, we restrict our attention hyper-cubes, $\ln
n_\ell=\nu$ for all $\ell$, and then
\begin{equation}
\label{Ad:sol}
S=
\int_{-\ii\infty}^{\ii\infty}\frac{dq_1}{2\pi \ii}\cdots
\int_{-\ii\infty}^{\ii\infty}\frac{dq_{d-1}}{2\pi \ii}\,
\Phi({\bf q})\,e^{\nu \phi({\bf q})}\,.
\end{equation}
Here, we introduced the shorthand notation \hbox{${\bf q}=(q_1,\ldots,
  q_{d-1})$} and
\begin{equation}
\phi({\bf q}) = \sum_{\ell=1}^{d-1} q_\ell - 1 + \frac{2^d}{\prod_{\ell=1}^{d-1} (q_\ell+1)}\,.
\end{equation}
The saddle point is ${\bf q}_*=(1,\ldots, 1)$, and at this point, it
is easy to show that $\phi({\bf q}_*)=d$ and $\Phi({\bf
  q}_*)=(2^d-1)/2^{d-1}$. We tacitly do not display the function
$\Phi({\bf q})$ because only its value at saddle point is needed.

To evaluate the integral \eqref{Ad:sol}, 
expand $\phi({\bf q})$ near the saddle point using 
\hbox{$q_\ell = 1 +  \ii u_\ell/\sqrt{\nu}$}, and then,  
\begin{equation}
\label{phi-ud}
\phi ({\bf u})\simeq d -  \frac{U({\bf u})}{\nu}, \quad U({\bf u}) =\frac{1}{2}
\sum_{a=1}^{d-1}\sum_{b=1}^a u_a u_b\,.
\end{equation}
The Gaussian integral can be computed in arbitrary dimension, 
\begin{equation}
\label{Gauss-d}
%\prod_{\ell=1}^{d-1}
\int_{-\infty}^\infty \!du_1
\cdots
\int_{-\infty}^\infty \!du_{d-1}
e^{-U({\bf u})} =
 \frac{(4\pi)^{(d-1)/2}}{\sqrt{d}}\,.
\end{equation}
The computation of the Gaussian integral relies on the fact that the
$(d-1)\times (d-1)$ matrix associated with the quadratic form $U({\bf
  u})$ has eigenvalues $\frac{1}{2}(d, 1, \dots, 1)$. The integral
\eqref{Gauss-d} completes the derivation of \eqref{F-d}.

\end{document}